\documentclass{ws-ijmpa}
\usepackage[super,compress]{cite}
\usepackage{graphicx}

\newcommand{\cN}{{\cal N}}
\newcommand{\Tr}{{\rm Tr\;}}
\newcommand{\hf}{\frac{1}{2}}
\newcommand{\qtr}{\frac{1}{4}}

\def\beq{\begin{equation}}
\def\eeq{\end{equation}}
\def\bea{\begin{eqnarray}}
\def\eea{\end{eqnarray}}
\def\nn{\nonumber}

\newcommand{\KD}{{K\"{a}hler-Dirac }}

\newcommand{\cA}{{\cal A}}
\newcommand{\cAb}{{\overline{\cal A}}}
\newcommand{\cD}{{\cal D}}
\newcommand{\cDb}{{\overline{\cal D}}}
\newcommand{\cF}{{\cal F}}
\newcommand{\cFb}{{\overline{\cal F}}}
\newcommand{\cM}{{\cal M}}
\newcommand{\cP}{{\cal P}}
\newcommand{\cO}{{\cal O}}
\newcommand{\cQ}{{\cal Q}}
\newcommand{\cU}{{\cal U}}
\newcommand{\cUb}{{\overline{\cal U}}}

\newcommand{\bn}{{\bf n}}
\newcommand{\hatbe}{{\widehat{e}}}

\begin{document}

\markboth{Anosh Joseph}
{Review of Lattice Supersymmetry and Gauge-Gravity Duality}

%
\catchline{}{}{}{}{}
%

\title{Review of Lattice Supersymmetry and Gauge-Gravity Duality\footnote{Invited review for the International Journal of Modern Physics A.}}

\author{ANOSH JOSEPH}

\address{John von Neumann Institute for Computing (NIC), \\
Deutsches Elektronen-Synchrotron DESY, \\
Platanenallee 6, D-15738 Zeuthen, GERMANY \\
Department of Applied Mathematics and Theoretical Physics (DAMTP), \\
University of Cambridge, Cambridge, CB3 0WA, UK \\
anosh.joseph@damtp.cam.ac.uk}

\maketitle

\begin{abstract}

We review the status of recent investigations on validating the gauge-gravity duality conjecture through numerical simulations of strongly coupled maximally supersymmetric thermal gauge theories. In the simplest setting, the gauge-gravity duality connects systems of D0-branes and black hole geometries at finite temperature to maximally supersymmetric gauged quantum mechanics at the same temperature. Recent simulations show that non-perturbative gauge theory results give excellent agreement with the quantum gravity predictions, thus proving strong evidence for the validity of the duality conjecture and more insight into quantum black holes and gravity. 

\end{abstract}

\ccode{PACS numbers:12.60.Jv; 11.25.Tq; 11.15.Ha; 04.70.Dy; 11.25.Uv; 04.60.Cf; 04.60.-m}

\tableofcontents

\section{Introduction}
\label{sec:intro}

The gauge-gravity duality conjecture, which is a generalization of the AdS/CFT correspondence, relates theories of quantum gravity living on curved spacetimes and quantum gauge field theories without gravity defined on the boundaries of such spacetimes. This conjecture provides very promising directions for the study of quantum gravity and black holes. It is possible to describe certain black holes in terms of the worldvolume theories of the D-branes that compose them. These theories are the maximally supersymmetric Yang-Mills (SYM) theories in various dimensions, taken in the 't Hooft limit and at finite temperature. They are strongly coupled in the regime in which they describe string theory black holes. Solving these thermal gauge theories at strong coupling would allow one to directly study the quantum properties of the dual black holes, including their thermodynamic aspects.    

The AdS/CFT correspondence relates maximally supersymmetric Yang-Mills theories with gauge group $SU(N)$, taken in the large $N$ 't Hooft limit to certain closed superstring theories in the near horizon region of $N$ coincident D-branes. The simplest case, which is the main focus of this review, is the one in which the gauge theory is the $SU(N)$ supersymmetric Yang-Mills quantum mechanics and its dual theory is type IIA string theory describing $N$ D0-branes.    

In this review, we focus on the recent developments in the gauge-gravity quality conjecture emphasizing the results from studying the strongly coupled supersymmetric gauge theories in one and two dimensions. To study these gauge theories at strong coupling directly we need to formulate them in a non-perturbative setting, and this can be achieved by regularizing them, in a consistent manner, on a spacetime lattice. We also discuss the lattice regularizations -- naive and manifestly supersymmetric -- of maximally supersymmetric Yang-Mills theories and the simulation algorithms needed to compute the interesting observables of the thermal gauge theories at strong coupling. 

This review is organized as follows. In section \ref{sec:qg-calcs} we briefly discuss the quantum gravity corrections to the internal energy of a black hole from low energy string theory calculations. We write down the internal energy of the quantum black hole as a function of its temperature with the anticipation that we can reproduce such a behavior from non-perturbative simulations of the dual thermal gauge theory. In sec. \ref{sec:BFSS-model} we write down the action of the sixteen supercharge Yang-Mills quantum mechanics theory -- the BFSS matrix model -- that takes part in the gauge-gravity duality. The BFSS model serves as the simplest setting in which we can probe the duality conjecture and related quantum gravity corrections. In sec. \ref{sec:twisted-version} we discuss a twisted version of the BFSS matrix model which shows compatibility with the lattice regularization by preserving a subset of the supersymmetry charges at finite lattice spacing. In sec. \ref{sec:latt-actions} we write down the lattice actions for maximally supersymmetric Yang-Mills theories that would allow us to explore the strong coupling regimes of these theories. In sec. \ref{sec:naive-latt-action} we write down a naive lattice action for the sixteen supercharge gauged quantum mechanics while in sec. \ref{sec:susy-latt-action} we write down a manifestly supersymmetric lattice action of the same theory. In sec. \ref{sec:sim-methods} we provide details of the simulation algorithms used to probe the strongly coupled regimes of these theories. We provide results from the recent lattice investigations in sec. \ref{sec:latt-results}, which support the predictions on the quantum gravity black holes from the gravity side of the duality. In sec. \ref{sec:BMN-model} we discuss a one-parameter deformation of the BFSS matrix model, known as the BMN model. This model is particularly interesting from the simulation point of view as it does not suffer from the thermal divergence of the partition function at smaller number of colors. In sec. \ref{sec:2d-16-theory} we provide a brief outlook on the supergravity system with D1-branes and associated black hole - black string phase transitions. In sec. \ref{sec:susy-lattice-2D} we discuss supersymmetric lattices for the two-dimensional maximally supersymmetric Yang-Mills. In sec. \ref{sec:results-GL} we provide results on the Gregory-Laflamme phase transitions from lattice simulations of the strongly coupled gauge theories. We conclude in sec. \ref{sec:future-directions} with the current status on validating the duality conjecture from the gauge theory side of the correspondence and possible future directions.

\section{D0-branes and Quantum Gravity Corrections}
\label{sec:qg-calcs}

String theory contains solitonic objects known as D-branes. In general, in a full string theory, there are modes that propagate in the bulk and also modes that propagate on the solitons. It is possible to define a limit of the full theory in which the bulk modes decouple from the modes that live on the solitons. This is typically the low energy limit of the theory, and in this limit, the D-brane theory becomes a supersymmetric Yang-Mills (SYM) theory. Since D-branes can carry mass and charge we can find supergravity solutions carrying the same mass and charge. Supergravity solutions are valid as long as the curvatures are locally small compared to the string scale $l_s$. When we have a system with large number $N$ of D-branes the curvature is small and we can trust the supergravity solution even at sub-stringy distances. Supergravity solutions corresponding to (0+1)-dimensional SYM theory, which we are interested in this review, are black D0-brane solutions. 

We study D0-branes in the field theory limit\cite{Maldacena:1997kk,Sen:1997we,Seiberg:1997ad}
\beq
\label{eq:ft-limit}
g^2_{YM} = \frac{1}{4\pi^2} \frac{g_s}{\alpha'^{\frac{3}{2}}} = {\rm fixed}, \quad \alpha' = l_s^2 \to 0,
\eeq
where $g_s$ is the string coupling and $g_{YM}$ the Yang-Mills coupling. We keep the energies fixed when we take the above limit. In our case it implies that the theory decouples from the bulk since the ten-dimensional Newton constant goes to zero. It also suppresses the higher order $\alpha'$ corrections\footnote{They come from the effects due to the strings having finite length.} in the action of the theory.   

When we take the limit given in Eq. \eqref{eq:ft-limit} we are interested in the finite energy configurations in the field theory. This corresponds to finite Higgs expectation values for some  fields of the theory. We are therefore considering the limit
\beq
U \equiv \frac{r}{\alpha'} = {\rm fixed}, \quad \alpha' \to 0,
\eeq
with $r$ representing some position at which the D0-brane is sitting, and it gives a Higgs expectation value $U$ to some fields of the theory. 

We will also consider near extremal configurations which correspond to the decoupled field theories at finite temperature. On the supergravity side we start from a near extremal black D0-brane solution and we take the limit Eq. \eqref{eq:ft-limit} keeping the energy density of the brane finite. In this limit the metric near the horizon has the following form\cite{Horowitz:1991cd,Gibbons:1987ps}
\bea
\label{eq:metric}
ds^2 &=& \alpha' \left(- \frac{ U^{\frac{7}{2}} }{ 2 \pi \sqrt{240 \pi^5 \lambda} } ( -f dt^2 ) + 2 \pi \sqrt{240 \pi^5 \lambda} \left( U ^{-\frac{7}{2}} \frac{dU^2}{f} + U^{-\frac{3}{2}} d \Omega_8^2 \right) \right),~~
\eea
with the function
\beq
f(U) = 1 - \frac{U_0}{U},
\eeq
and $d \Omega_8^2$ representing the line element of an eight-dimensional unit sphere. The event horizon of the black hole is located at $U = U_0$. The parameters $\lambda$ and $U_0$ are related to the mass and charge of the supergravity black hole.  

The black hole described above appears as a quantum system with temperature $T$ for a distant observer. It undergoes thermal evaporation through black body radiation. The Hawking temperature characterizing this is obtained by a standard calculation
\beq
\frac{T}{\lambda^{\frac{1}{3}}} = \frac{7}{16 \pi^3 \sqrt{15 \pi}} \left(\frac{U_0}{\lambda^{\frac{1}{3}}}\right)^{\frac{5}{2}}.
\eeq

The Bekenstein-Hawking entropy of the black hole is evaluated by the area $A$ of the horizon in the Einstein frame as\cite{Klebanov:1996un}
\beq
\frac{S}{N^2} = \frac{1}{N^2} \frac{A}{4 G_N} = \left( \frac{ 2^{26} 15^2 \pi^{14} }{ 7^{14} } \right)^{\frac{1}{5}} \left( \frac{T}{ \lambda^{\frac{1}{3}} }\right)^{\frac{9}{5}}. 
\eeq
  
The gauge-gravity duality tells us that $T$ and $S$ should be identified with the temperature and entropy of the dual gauge field theory. We can make use of the first law of thermodynamics, $dE = T dS$, to obtain the internal energy $E \equiv E(T)$ of the black hole. Normalizing by $\lambda^{\frac{1}{3}}$ we have  
\beq
\label{eq:energy-bh}
\frac{E}{N^2} = \frac{9}{14} \left( \frac{ 2^{26} 15^2 \pi^{14} }{ 7^{14} }\right)^{\frac{1}{5}} T^{\frac{14}{5}} = 7.41 ~T^{2.8},
\eeq
where $E$ and $T$ are now dimensionless variables.

It is possible to add quantum corrections to Eq. \eqref{eq:energy-bh}. The back reaction of the Hawking radiation can contribute as non-local quantum corrections to the supergravity action. The magnitude of such corrections can be estimated by assuming the Stefan-Boltzmann law and using the black hole geometry given in Eq. \eqref{eq:metric}. The calculations show that the energy loss due to Hawking radiation is suppressed by a factor of $O(l_s^{14})$ and it indeed vanishes in the low energy limit, $l_s \to 0$, considered here.

One could also add another correction to Eq. \eqref{eq:energy-bh} due to the quantum gravity effects arising from short distances. One can obtain such a correction by calculating perturbatively the scattering amplitude involving four gravitons as asymptotic states in superstring theory. This amplitude is coming from an intermediate process of strings being pair-created and pair-annihilated near the horizon. Such quantum corrections can be implemented into the supergravity action by adding quadratic terms of the Riemann tensor at the leading order as shown by Gross and Witten\cite{Gross:1986iv}. Including such quantum corrections one obtains the metric near the horizon as 
\bea
\label{eq:metric-near-horr}
ds^2 &=& \alpha' \left( - \frac{\sqrt{H_2} F_1}{H_1} dt^2 + \frac{\sqrt{H_2}}{F_1} dU^2 + \sqrt{H_2} U^2 d\Omega_8^2 \right),
\eea
where
\bea
H_i(U) &=& \frac{ 240 \pi^5 \lambda }{ U_0^7 } \left( \frac{U_0^7}{U^7} + \frac{ \pi^6 \lambda^2 }{ 2^7 3^2 N^2 U_0^6 } h_i (U) \right),
\eea
\bea
\label{eq:F1U}
F_1(U) &=& 1 - \frac{ U_0^7 }{ U^7 } + \frac{ \pi^6 \lambda^2 }{ 2^7 3^2 N^2 U_0^6 } f_1 (U),
\eea
and $h_i$, $i =1,2$ and $f_1$ are functions of $U/U_0$. The functions $H_{1,2}(U)$ and $F_1(U)$ can be determined uniquely to the order $1/N^2$ by imposing appropriate boundary conditions as shown by Hyakutake\cite{Hyakutake:2013vwa}. The metric given in Eq. \eqref{eq:metric-near-horr} reduces to the one in Eq. \eqref{eq:metric} when $N \to \infty$, which corresponds to the limit of classical gravity. It is seen from Eq. \eqref{eq:F1U} that the position of the event horizon is slightly shifted inward due to quantum corrections as
\beq
U = U_0 - \frac{ \pi^6 \lambda^2 }{ 2^7 3^2 N^2 U_0^6 } f_1 (U_0),
\eeq
where $f_1(U_0)$ is some positive constant. 

In the modified background geometry given in Eq. \eqref{eq:metric-near-horr} one can proceed to calculate the Hawking temperature in a standard manner. We need to use Wald's entropy formula, which is applicable to the present case with higher curvature terms, in order to evaluate the entropy of the quantum black hole. The temperature and entropy of the black hole can be identified with the corresponding ones in the dual gauge theory with the contributions coming from the $1/N^2$ corrections. The internal energy $E$ in this case can also be obtained directly from the first law of thermodynamics. Thus we arrive at 
\beq
\label{eq:energy-bh2}
\frac{E}{N^2} = 7.41 ~T^{2.8} - 5.77 ~T^{0.4}\frac{1}{N^2},
\eeq
where the second term represents the quantum gravity corrections at the leading order.

We have ignored the $\alpha'$ corrections in the above analysis. One can include these corrections to Eq. \eqref{eq:energy-bh2} as has been done in Ref. [\refcite{Hanada:2008ez}] at $N = \infty$. This gives the following result for the internal energy of a quantum black hole
\bea
\label{eq:alpha-prime-corrected}
\frac{1}{N^2} E_{\rm gravity} &=& \left( 7.41 ~T^{2.8} + A ~T^{4.6} + \cdots \right) \nn \\
&&\quad \quad \quad \quad + \left( -5.77 ~T^{0.4} + B ~T^{2.2} + \cdots \right) \frac{1}{N^2} + O \left( \frac{1}{N^4} \right),
\eea
where the constants  $A$ and $B$ are not yet known. We have denoted the internal energy by $E_{\rm gravity}$ in the above expression since this result is coming from calculations on the gravity side of the duality. We will denote the internal energy obtained from the gauge theory side of the correspondence as $E_{\rm gauge}$. The power of $T$ for each term in the above result can be determined from dimensional analysis using some known results in superstring theory\cite{Green:2006gt}.
  
\section{The BFSS Matrix Model}
\label{sec:BFSS-model}

The BFSS matrix model\cite{Banks:1996vh} is the maximally supersymmetric matrix quantum mechanics, which is obtained by dimensionally reducing four-dimensional maximally supersymmetric Yang-Mills theory to one dimension. This model is conjectured to describe a discrete light-cone quantization of M-theory\cite{Banks:1996vh} on a circle. It also describes, with suitably re-identified parameters, the low energy dynamics of a collection of $N$ D0-branes of type IIA superstring theory\cite{Witten:1995ex}. This theory also plays the role as a matrix regularization of the light-cone action for the 11-dimensional super-membrane. (See [\refcite{Taylor:2001vb}] for a review.)

The theory contains a $U(N)$ gauge field $A_0$, nine scalars $X_i$ ($i = 1, \cdots, 9$), which are $N \times N$ hermitian matrices, and 16 fermions $\psi_\alpha$ ($\alpha = 1, \cdots, 16$). The continuum Euclidean action of the theory has the form 
\bea
\label{eq:action-BFSS-eucl}
S_E &=& \frac{N}{2\lambda} \int d\tau~ \Tr \Big\{ \hf (D_\tau X_i)^2 - \qtr [X_i, X_j]^2 + \hf \psi_\alpha D_\tau \psi_\alpha \nn \\
&& \qquad \qquad \qquad \qquad + \hf \psi_\alpha (\gamma_i)_{\alpha \beta} [X_i, \psi_\beta] \Big\},
\eea 
where $\lambda \equiv N g^2_{YM}$ is the 't Hooft coupling and $\gamma_i$ are real symmetric matrices satisfying the nine-dimensional Clifford algebra, $\{\gamma_i, \gamma_j\} = 2\delta_{ij}$. 

The covariant derivative $D_\tau$ is defined as 
\beq
D_\tau \cdot = \partial_\tau \cdot + i [A_0, ~\cdot~].
\eeq

The theory is invariant under the following set of supersymmetry transformations
\bea
\delta_\xi &=& \xi_\alpha \psi_\alpha, \nn \\
\label{eq:susy-trans}
\delta_\xi X_i &=& -i \xi_\alpha (\gamma_i)_{\alpha \beta} \psi_\beta, \\
\delta_\xi \psi_\alpha &=& i (\gamma_i)_{\alpha \beta} \xi_\beta D_\tau X_i - \hf (\Sigma_{ij})_{\alpha \beta} \xi_\beta [X_i, X_j], \nn
\eea
where 
\beq
\Sigma_{ij} = \frac{i}{2} [\gamma_i, \gamma_j]
\eeq
and $\xi_\alpha$ are sixteen fermionic parameters. The invariance of the action Eq. \eqref{eq:action-BFSS-eucl} under the set of transformations Eq. \eqref{eq:susy-trans} can be shown using the Fierz identity for the fermions and the Jacobi identity for the scalars. 

The supersymmetry charges $Q_\alpha$ of the theory are the generators of the infinitesimal transformations, 
\beq
\delta_\xi = \xi_\alpha Q_\alpha.
\eeq

In Sec. (\ref{sec:naive-latt-action}) we study a lattice version of this theory, which is obtained through a naive lattice regularization. In Sec. (\ref{sec:susy-latt-action}) we look at another lattice discretization of this theory, which is manifestly supersymmetric;  it preserves two of the sixteen supersymmetry charges on the lattice.

\subsection{Topological Twist of the BFSS Matrix Model}
\label{sec:twisted-version}

The BFSS matrix model can be expressed in a convenient form, which is compatible with lattice regularization, after relabeling its fields and supersymmetries. The lattice theory obtained through such a prescription is manifestly gauge-invariant and preserves two or eight supersymmetry charges exact, depending on the regularization scheme, at finite lattice spacing. This method of relabeling the fields and supercharges is known as {\it topological twisting}. The twisting process is done on its four-dimensional parent theory, the $\cN = 4$ SYM theory. The four-dimensional theory has $SO(4)$ Euclidean spacetime rotation group and $SO(6)$ internal rotation (R-symmetry) group. We introduce a new rotation group $SO(4)'$ by taking the diagonal subgroup
\beq
SO(4)' = {\rm diag~subgroup}~\left(SO(4)_{\rm Lorentz} \times SO(4)_R\right).
\eeq

The theory obtained this way is called twisted theory. The fields and supersymmetries of the original theory are then expressed in terms of the twisted Lorentz rotation group. 

We dimensionally reduce the four-dimensional twisted theory to one dimension to obtain a twisted version of the BFSS matrix model. This model can be easily transported on to the lattice by preserving gauge symmetry and a part of the supersymmetry. In this section we discuss the twisted theory that leads to a lattice action proposed by Sugino. In Ref. [\refcite{Kadoh:2012bg,Kadoh:2014hsa,Kadoh:2015mka}] Kadoh and Kamata performed a series of lattice simulations of this theory and examined the gauge-gravity duality in this system. We note that there also exist a complementary lattice discretization of the 16 supercharge theory, based on the idea of orbifolding, first proposed by Kaplan and \"Unsal \cite{Kaplan:2005ta, Unsal:2006qp}. This construction preserves 8 supercharges at finite lattice spacing. It would be interesting to carry out a detailed nonperturbative numerical study in the system prescribed by such a regularization scheme.

Let us briefly discuss the twisted version of the BFSS matrix model. The transformation laws of the fields and supercharges under the twisted Lorentz group are different from those of the untwisted theory. We can construct two linear combinations of supercharges from the original sixteen supercharges, which we label as $Q_\pm$. These two supercharges, which we call as twisted supercharges, are scalar fermions and nilpotent up to a gauge rotation. After twisting, we can express the action of the theory in a topological $Q_\pm$-exact form. Then the $Q_\pm$-invariance of the action follows from the nilpotency of $Q_\pm$.

We change the field variables of the original theory to those of the twisted theory in order to define the $Q_\pm$-exact action directly from Eq. \eqref{eq:action-BFSS-eucl}. The bosonic twisted fields are denoted by $A_\mu, B_i, C$ and $ \phi_\pm$, with $\mu = 0, 1, 2, 3$ and $i = 0, 1, 2$, and the the fermionic twisted fields are denoted by $\eta_\pm, \psi_{\pm \mu}$ and $\chi_{\pm i}$. Under the four-dimensional twisted Lorentz transformations the fields $A_\mu$ and $\psi_{\pm \mu}$ transform as vectors, the fields $C, \phi_\pm$ and $\eta_\pm$ remain unchanged (they transform as scalars), and the other bosons and fermions of the theory transform as self-dual tensors. The fields $B_i$ and $\chi_{\pm i}$ are three independent components of the bosonic and fermionic tensors, respectively. In one dimension we associate the indices $\mu, i$ to the names of the fields. 

The two nilpotent twisted supercharges $Q_\pm$ can be constructed as linear combinations of the supersymmetry charges of the original theory in the following way
\bea
Q_+ &=& \frac{1}{\sqrt{2}} (Q_5 + i Q_{13}), \\
Q_- &=& \frac{1}{\sqrt{2}} (Q_1 + i Q_9),
\eea
where $Q_i$ denotes the $i$-th component of the sixteen-component supersymmetry charge.

The twisted fields in the bosonic sector are related to the untwisted fields the following way
\begin{align}
X_\mu  &= A_\mu,~~\mu = 1,2,3   & \\
X_4  &= -B_2   &   X_5  &= B_1 \\
X_6  &= -B_0   &   X_7  &= \hf C \\
X_8 &= \hf (\phi_+ - \phi_-)       &   X_9 &= \frac{i}{2} (\phi_+ + \phi_-)
\end{align}
with $A_0$ unchanged. The fields $A_\mu, B_i$ and $C$ are hermitian and $(\phi_+)^\dagger = - \phi_-$.

The fermionic twisted fields $\eta_\pm, \psi_{\pm \mu}$ and $\chi_{\pm i}$ are related to the components of the untwisted sixteen-component fermion field $\psi_\alpha$ the following way 
\beq
\left( \begin{array}{c}
\psi_1 \\
\psi_2 \\
\psi_3 \\
\psi_4 \\
\psi_5 \\
\psi_6 \\
\psi_7 \\
\psi_8 \\
\psi_9 \\
\psi_{10} \\
\psi_{11} \\
\psi_{12} \\
\psi_{13} \\
\psi_{14} \\
\psi_{15} \\
\psi_{16} \end{array} \right) = \frac{1}{\sqrt{2}}\left( \begin{array}{c}
\psi_{-0} + \frac{i}{2} \eta_+ \\
\psi_{-1} - i \chi_{+2} \\
\psi_{-2} + i \chi_{+1} \\
\psi_{-3} - i \chi_{+0} \\
\psi_{+0} + \frac{i}{2} \eta_- \\
\psi_{+1} + i \chi_{-2} \\
\psi_{+2} - i \chi_{-1} \\
\psi_{+3} + i \chi_{-0} \\
-i(\psi_{-0} - \frac{i}{2} \eta_+) \\
-i(\psi_{-1} + i \chi_{+2}) \\
-i(\psi_{-2} - i \chi_{+1}) \\
-i(\psi_{-3} + i \chi_{+0}) \\
-i(\psi_{+0} - \frac{i}{2} \eta_-) \\
-i(\psi_{+1} - i \chi_{-2}) \\
-i(\psi_{+2} + i \chi_{-1}) \\
-i(\psi_{+3} - i \chi_{-0}) \end{array} \right).
\eeq
We use appropriate set of gamma matrices (see Ref. [\refcite{Kadoh:2015mka}]) to connect the two sets (twisted and untwisted) of the field components. 

It is straightforward to obtain the $Q_\pm$-transformation laws on the twisted fields. They are given by
\bea
Q_\pm A_\mu &=& \psi_{\pm \mu} \\
Q_\pm \psi_{\pm \mu} &=& -i D_\mu \phi_\pm \\
Q_\pm B_i &=& \chi_{\pm i} \\
Q_\pm \chi_{\pm i} &=& [B_i, \phi_\pm] \\
Q_\pm C &=& \eta_\pm \\
Q_\pm \eta_\pm &=& [C, \phi_\pm] \\
Q_\pm \phi_\mp &=& \eta_\mp \\
Q_\pm \eta_\mp &=& [\phi_\mp, \phi_\pm] \\
Q_\pm \phi_\pm &=& 0 \\ 
Q_\pm \chi_{\mp i} &=& \hf [C, B_i] \pm H_i \\
Q_\pm \psi_{\mp \mu} &=& \frac{i}{2} D_\mu C \pm \widetilde{H}_\mu \\
Q_\pm H_i &=& \pm \left( [\chi_{\mp i}, \phi_\pm] + \hf [\chi_{\pm i}, C] + \hf [B_i, \eta_\pm] \right) \\
Q_\pm \widetilde{H}_\mu &=& \pm \left( [\psi_{\mp \mu}, \phi_\pm] + \hf [\chi_{\pm \mu}, C] - \frac{i}{2} D_\mu \eta_\pm] \right)
\eea
where $D_0$ is the covariant derivative and $D_\mu$ with $\mu = 1, 2, 3$, are the commutators $D_\mu \cdot = i [A_\mu, ~\cdot~]$. We have used the four index notation for $D_\mu$ ($\mu = 0, 1, 2, 3$) since the twisted theory is originally defined in four dimensions. We have also introduced seven auxiliary fields $H_i$ and $\widetilde{H}_\mu$ to make the $Q_\pm$ transformations satisfy the following relations
\bea
Q_\pm^2 &=& i \delta_{\phi_\pm}, \\
\{ Q_+, Q_- \} &=& -i \delta_C,
\eea 
where the right-hand sides correspond to gauge rotations with field dependent gauge parameters. 

After relabeling the variables of the original theory we write down the action of the twisted theory in a $Q_\pm$-exact form
\beq
\label{eq:twisted-action}
S = \frac{N}{2 \lambda}  Q_+ Q_- ~\Lambda,
\eeq
where the ``gauge fermion" $\Lambda$ is 
\bea
\Lambda &=& \int d\tau~ \Tr \Big\{ -2i B_i \Big(F_{i3} + \hf \epsilon_{ijk} F_{jk}\Big) + \frac{1}{3} \epsilon_{ijk} B_i [B_j, B_k] \nn \\
&& \qquad \qquad \qquad \qquad -\psi_{+\mu} \psi_{-\mu} + \chi_{+i} \chi_{-i} - \qtr \eta_+ \eta_- \Big\},
\eea
with $F_{01} = D_0 A_i$ and $F_{ij} = i [A_i, A_j]$ with $i, j = 1,2,3$.

We see that the twisted action is trivially invariant under the $Q_\pm$-transformations, without invoking the Leibniz rule, owing to the nilpotency of the twisted supersymmetry charges. The other fourteen supersymmetry charges and the global $SO(9)$ symmetry are also exact symmetries of the continuum twisted action since we have only performed a change of variables of the original theory to obtain the twisted theory.

\section{Lattice Actions}
\label{sec:latt-actions}

There are two ways to discretize the sixteen supercharge quantum mechanics model on the lattice. One is a naive discretization of the continuum theory and the other is a manifestly supersymmetric lattice action. The naive lattice action follows the standard rules of discretization of a gauge theory. The resulting fermion determinant can be complex in general at finite lattice spacing, which can make the simulation problematic. While the naive lattice action preserves no supersymmetry, we can argue that since the quantum mechanics is free from UV divergences given that the gauge and global symmetries are preserved by a UV regulator, all the supersymmetries will be restored in the continuum limit. The twisted version of the theory is manifestly supersymmetric and gauge-invariant on the lattice. It preserves two supersymmetry charges exact at finite lattice spacing while the other fourteen supercharges are broken on the lattice. Supersymmetry is broken softly in both the lattice constructions by thermal boundary conditions and also by the addition of mass terms for the scalars to stabilize the simulations. 

We discretize the theory on a Euclidean time circle consisting of $M$ number of sites, $n = \{0, \cdots, M - 1\}$. Thermal boundary conditions correspond to taking the fermions anti-periodic on the Euclidean time circle. The temperature of the system is $T = 1/\beta$, with $\beta = aM$ and $a$ denotes the lattice spacing.

\subsection{Naive Lattice Action}
\label{sec:naive-latt-action}

The continuum Euclidean path integral of the sixteen supercharge theory takes the following from after the fermions are integrated out
\beq
Z = \int DA DX ~{\rm Pf} (\cM) e^{-S_B},
\eeq
where ${\rm Pf}(\cM)$ denotes the Pfaffian of the fermion operator $\cM$. The bosonic action is
\beq
S_B = \frac{N}{2\lambda} \Tr \oint^\beta d\tau \left\{ \hf (D_\tau X_i)^2 - \qtr [X_i, X_j]^2 \right\},
\eeq
and the fermion operator
\beq
\cM = \gamma_\tau D_\tau - \gamma_i [X_i, ~\cdot~ ].
\eeq

We take $\gamma_\tau, \gamma_i$ in the Euclidean representation of the Lorentzian Majorana-Weyl gamma matrices, obeying $\{\gamma_i, \gamma_j\} = 2\delta_{ij}$ for index $\nu = \{\tau, i\}$. The Pfaffian is in general complex\cite{Krauth:1998xh}, giving rise to a potential sign problem. It is thus important, in principle, to include the phase of the Pfaffian in the Monte Carlo simulation. If the phase of the Pfaffian is very nearly real and positive we can ignore the phase and proceed with phase quenched simulations\footnote{See Refs. [\refcite{Catterall:2011aa,Mehta:2011ud,Galvez:2012sv}] for a recent work on the sign problem in 4 and 16 supercharge Yang-Mills theories in two dimensions.}.

We also need to add mass terms to the lattice action in order to regulate the flat directions associated with the scalars in the theory. The classical potential, $\sim \Tr ([X_i, X_j]^2)$, has flat directions, that is, any set of matrices which are mutually commuting have zero energy. These flat directions can be removed by adding mass terms of the form
\beq
S_M = \frac{N}{2\lambda} \oint^\beta d\tau \sum_{i = 1}^9 \Tr \hf \mu^2 \left( X_i^2 \right),
\eeq 
with $\mu$ a mass parameter. In general one simulates the theory at various values of the mass parameter and then extrapolates to vanishing mass limit. 

We discretize the continuum model as,
\beq
S_B = \frac{N}{2\lambda} \frac{M^3}{\beta^3} \sum_{n = 0}^{M-1} \Tr \Big( \hf (D_- X_i)_n - [X_{i,n}, X_{j,n}]^2 + \hf \mu_0^2 X_{i,n}^2 \Big),
\eeq
and
\beq
\cM_{mn} = \left( \begin{array}{cc}
0 & (D_-)_{mn} \\
-(D_-)_{nm} & 0 \end{array} \right) - \gamma^i [X_{i,n}, \cdot~]~{\bf I}_{mn}, 
\eeq
where we have rescaled the fields $X_{i,n}$ and $\psi_{\alpha,n}$ by powers of the lattice spacing to render them dimensionless. The dimensionless lattice mass regulator $\mu_0$ is related to the continuum mass as 
\beq
\mu_0 = \beta \mu,
\eeq
and it is this dimensionless $\mu_0$ that parameterizes how much of the weakly-coupled divergent region is allowed by the mass regulator\cite{Catterall:2009xn}. 

We define the lattice covariant difference operator the following way 
\beq
(D_- W_i)_n = W_{i,n} - U_n W_{i, n-1} U^\dagger_n,
\eeq 
and we have introduced a Wilson gauge link field $U_n$. Note that the fermionic operator $\cM$ is free of doublers and manifestly antisymmetric on the lattice. We have chosen a twisted Euclidean representation for the gamma matrices in order to obtain this antisymmetric fermion operator after taking finite differences. We have taken a representation of gamma matrices where
\beq
\gamma_\tau = i \left( \begin{array}{cc}
0 & {\bf I}_8 \\
{\bf I}_8 & 0 \end{array} \right).
\eeq

In a super-renormalizable theory only a finite number of divergences can arise in the lattice regularized theory. Such divergences may induce a RG flow away from the maximally supersymmetric continuum limit of our interest. However, one would naively expect that for quantum mechanics there will be no divergences. Thus we expect that the lattice action is finite and hence will flow without fine-tuning to the correct maximally supersymmetric continuum theory as the lattice spacing is decreased. In Ref. [\refcite{Catterall:2008yz}] this was tested by simulating the theory with periodic boundary conditions and testing whether the path integral, the Witten index, was dependent of $\beta$ in the continuum limit. 

\subsection{Manifestly Supersymmetric Lattice Action}
\label{sec:susy-latt-action}

It is straightforward to discretize the topologically twisted theory given in Eq. \eqref{eq:twisted-action}. We place the adjoint scalars and fermions on the sites and the lattice gauge field $U_n \in SU(N)$ on the links. The covariant difference operator on the lattice acts the following way on a generic field $\varphi_n$
\beq
D^+ \varphi_n = U_n \varphi_{n+1} U^{-1}_n - \varphi_n
\eeq
and it transforms as $\varphi_n$ itself. 

By replacing the integral and the covariant derivative in the twisted action Eq. \eqref{eq:twisted-action} with the summation over the lattice sites and the covariant difference operator, respectively, we obtain the following lattice action (We call this the Sugino lattice action\cite{Sugino:2004uv} as it is the dimensional reduction of the four-dimensional lattice theory constructed by Sugino.)
\bea
\label{eq:latt-susy-exact-action}
S &=& Q_+ Q_- \frac{N}{2 \lambda_0}~\Lambda,
\eea
where
\bea
\Lambda &=& \sum_{n=0}^{M-1} \Tr \Big\{ -2i B_i \Big(F_{i3} + \hf \epsilon_{ijk} F_{jk}\Big) + \frac{1}{3} \epsilon_{ijk} B_i [B_j, B_k] \nn \\
&& \qquad \qquad \qquad \qquad -\psi_{+\mu} \psi_{-\mu} + \chi_{+i} \chi_{-i} - \qtr \eta_+ \eta_- \Big\},
\eea
with
\beq
F_{0i} = D^+ A_i~~{\rm and}~~F_{ij} = i [A_i, A_j],~~i, j = 1, 2, 3,
\eeq
and we have denoted 
\beq
\lambda_0 = \lambda a^3
\eeq
as the dimensionless 't Hooft coupling. The continuum limit of this theory is obtained by taking the limit $\lambda_0 \to 0$ with $\lambda$ fixed.

The $Q_\pm$ transformations on the twisted lattice fields take the following form 
\bea
Q_\pm U &=& i \psi_{\pm 0} U \\
Q_\pm A_\mu &=& \psi_{\pm \mu} \\
Q_\pm \psi_{\pm \mu} &=& -i D_\mu \phi_\pm + i \delta_{\mu 0} \psi_{\pm 0} \psi_{\pm 0} \\
Q_\pm B_i &=& \chi_{\pm i} \\
Q_\pm \chi_{\pm i} &=& [B_i, \phi_\pm] \\
Q_\pm C &=& \eta_\pm \\
Q_\pm \eta_\pm &=& [C, \phi_\pm] \\
Q_\pm \phi_\mp &=& \eta_\mp \\
Q_\pm \eta_\mp &=& [\phi_\mp, \phi_\pm] \\
Q_\pm \phi_\pm &=& 0 \\ 
Q_\pm \chi_{\mp i} &=& \hf [C, B_i] \pm H_i \\
Q_\pm \psi_{\mp \mu} &=& \frac{i}{2} D_\mu C \pm \widetilde{H}_\mu + \frac{i}{2} \delta_{\mu 0} \{\psi_{+0}, \psi_{-0}\} \\
Q_\pm H_i &=& \pm \left( [\chi_{\mp i}, \phi_\pm] + \hf [\chi_{\pm i}, C] + \hf [B_i, \eta_\pm] \right) \\
Q_\pm \widetilde{H}_\mu &=& \pm \left( [\psi_{\mp \mu}, \phi_\pm] + \hf [\chi_{\pm \mu}, C] - \frac{i}{2} D_\mu \eta_\pm] \right) \nn \\
&&\pm \delta_{\mu 0} \left(\qtr [\psi_{\pm 0}, D_0 C \pm 2i H + \{ \psi_{+ 0}, \psi_{- 0} \} ] + \hf [\psi_{\mp 0}, D_0 \phi_\pm] \right)
\eea
They are similar to the corresponding continuum transformations with the exception of the fields with index $\mu = 0$. The terms containing $\delta_{\mu 0}$ are higher order corrections. 

The supersymmetry charges $Q_\pm$ satisfy the following set of relations on the lattice\cite{Sugino:2004uv}
\bea
Q_\pm^2 &=& i \delta_{\phi_\pm}, \\
\{ Q_+, Q_- \} &=& i \delta_C,
\eea
which resembles the corresponding set of relations in the continuum. Such a resemblance does not come as a surprise if we recall the BRST transformations. In the BRST transformations, once the transformation rule for a lower dimensional field is given, then that of the higher dimensional field is uniquely fixed from the nilpotency property. The same can be applied on the lattice, once the $Q_\pm$-transformation laws of the lower dimensional fields, $Q_\pm U, Q_\pm B_i,~etc.$, are given, then the others, $Q_\pm \psi, Q_\pm \chi_\pm,~etc.$, are uniquely determined from the above nilpotency equations.

The lattice action has $Q_\pm$-invariance and gauge-invariance. We note that the remaining supercharges and the $SO(9)$ global rotational symmetry are not preserved on the lattice. The lattice action has, at least, an $SU(2)$ global symmetry corresponding to the interchange of $Q_+$ and $Q_-$. The lattice $Q_\pm$ transformations reproduce the corresponding set of continuum ones and the lattice action reproduces the correct continuum action in the naive continuum limit. 

After performing $Q_\pm$-variations on the lattice action Eq. \eqref{eq:latt-susy-exact-action} we find that it has a four-Fermi term,
\beq
S_{\rm 4-Fermi} = \frac{N}{2 \lambda_0} \sum_n \Tr \left(-\qtr \{\psi_{+0}, \psi_{-0}\}^2 \right),
\eeq
which is of the order of the cut-off. The four-Fermi interaction term is not suitable for numerical simulations. One could express the four-Fermi term as a fermion bilinear form, as in the NJL model, by introducing an auxiliary scalar field $\sigma$. The four-Fermi term becomes 
\beq
S_{\rm 4-Fermi} = \frac{N}{2 \lambda_0} \sum_n \Tr \left( \sigma^2 + \psi_{+0} [\sigma, \psi_{-0}] \right).
\eeq
It reproduces the original four-Fermi piece once $\sigma$ is integrated out.

\section{Lattice Simulations of Sixteen Supercharge Yang-Mills}
\label{sec:sim-methods}

The maximally supersymmetric Yang-Mills theory can be simulated on the lattice using rational hybrid Monte Carlo (RHMC) algorithm introduced in Ref. [\refcite{Clark:2006wq}]. We are interested in numerically computing operator expectation values $\langle \cO \rangle$ through a discretized version of the path integral
\beq
\langle \cO \rangle = \frac{1}{Z} \int D\Phi D\Psi~ \cO ~e^{-S[\Phi, \Psi]},
\eeq
where the partition function is
\beq
Z = \int D\Phi D\Psi~ \cO ~e^{-S[\Phi, \Psi]},
\eeq
and $\Phi$ and $\Psi$ denote, respectively the set of bosons and fermions in the theory. The discretized path integral is evaluated through the importance sampling Monte Carlo algorithm, which we briefly discuss below.  

We first integrate out the fermions in the theory and this leads to a Pfaffian. The partition function becomes
\beq
Z = \int D\Phi~ {\rm Pf}(\cM(\Phi))~ e^{-S_B[\Phi]}.
\eeq

The Pfaffian of the theory can be complex in general,
\beq
{\rm Pf}(\cM) = \left| {\rm Pf}(\cM) \right| e^{i \alpha},
\eeq
with a phase $\alpha$ and it can affect the stability of numerical simulations. Investigations on the sixteen supercharge lattice theory show that this model does not suffer from a severe sign problem -- the Pfaffian is very nearly real and positive\cite{Catterall:2009xn,Filev:2015hia}. Thus we can ignore the phase of the Pfaffian and use the absolute value
\beq
\left| {\rm Pf} (\cM) \right| = \left| \det (\cM) \right|^\hf = \det \left(\cM^\dagger \cM \right)^\qtr.
\eeq

The next step is to introduce a set of bosonic pseudo-fermions $\xi$ and express the absolute value of the Pfaffian
\beq
\left| {\rm Pf} (\cM) \right| = \int D \xi^\dagger D \xi \exp\left( - \xi^\dagger (\cM^\dagger \cM)^{-\qtr} \xi \right), 
\eeq
where $\xi$ is a $16(N^2 -1)$-component vector (for $SU(N)$ gauge group) on each lattice site. 

At each site $n$ the pseudo-fermion field is set to 
\beq
\xi_n = (\cM^\dagger \cM)^{\frac{1}{8}} \eta_n,
\eeq
with $\eta_n$ being $16(N^2-1)$-component random vectors drawn from a Gaussian distribution. 

In order to simulate the theory we need to approximate the rational exponent of the squared fermion operator $\cM^\dagger \cM$ with a partial sum
\beq
(\cM^\dagger \cM)^{-\qtr} = \alpha_0 + \sum_{i=1}^P \frac{\alpha_i}{\cM^\dagger \cM + \beta_i},
\eeq
where the degree $P$, and the coefficients $\alpha_i, \beta_i$ known as amplitudes and shifts, respectively depend on the range of the eigenvalues of $\cM^\dagger \cM$ and the accuracy of the approximation. We can compute the optimal shifts and amplitudes offline, through the Remez algorithm\cite{AlgRemez:2005}, by minimizing the relative approximation error within a given spectral range $[\lambda_{\rm low}, \lambda_{\rm high}]$, with $\lambda_{\rm low} > 0$ of the squared fermion operator. We need to add more terms to the partial fraction approximation if we want to keep the errors negligible across larger spectral ranges and this in turn results in increased computational costs. We must demand, at a minimum, that the smallest and largest eigenvalues of $\cM^\dagger \cM$ lie within the spectral range we use, that is, $\lambda_{\rm low} < \lambda_{\rm min} \ll \lambda_{\rm max} < \lambda_{\rm high}$. We should also monitor the extremal eigenvalues of the squared fermion operator, like the Pfaffian, during the RHMC evolution.  
  
Following the standard procedure, we then introduce fictitious momenta $p_\Phi$, drawn from a Gaussian distribution, that are conjugate to the bosonic coordinates $\Phi$ and evolve the coupled system using molecular dynamics (MD) evolution along a trajectory of length $\tau$ in a fictitious simulation time $t$. The effective Hamiltonian has the form
\beq
H = \hf \sum_n \Tr \left(p_{\Phi_n}^2\right) + S_B(\Phi_n) + \xi_n^\dagger ( \cM^\dagger \cM ) \xi_n. 
\eeq

We keep the pseudo-fermion field fixed along the trajectory as it already has a proper distribution. Since the MD evolution involves inexact integration of Hamilton's equations, which does not conserve the effective Hamiltonian, we need to impose a criterion to accept or reject the new set of field configurations at the end of the trajectory. The new field configurations produced by the MD evolution are accepted or rejected using the Metropolis-Rosenbluth-Teller test, with a probability
\beq
p = {\rm min}(1, e^{-\Delta H}),
\eeq
where $\Delta H$ is the change in effective Hamiltonian after one MD evolution. The original configuration values are restored back if the new configurations are not accepted and the evolution is repeated from a fresh set of Gaussian distributions for the momenta. The test makes the algorithm exact by stochastically correcting for integration errors as long as the MD integration scheme we use is area preserving (symplectic) and reversible (symmetric). 

Each step of the MD evolution requires solving the following set of equations 
\bea
\frac{\delta \Phi}{\delta t} &=& \frac{\delta H}{\delta p_\Phi} = p_\Phi, \\
\frac{\delta p_\Phi}{\delta t} &=& - \frac{\delta H}{\delta \Phi} = - \frac{\delta S_B}{\delta \Phi} - \frac{\delta}{\delta \Phi} \xi^\dagger (\cM^\dagger \cM)^{-\qtr} \xi.
\eea

The gradients of the fields with respect to the pseudo-fermion action (the fermionic forces) have the form
\bea
\frac{\delta}{\delta \Phi} \xi^\dagger (\cM^\dagger \cM)^{-\qtr} \xi &=& \sum_{i=1}^P \alpha_i \Big[(\cM^\dagger \cM + \beta_i)^{-1} \xi\Big] \frac{\delta(\cM^\dagger \cM)}{\delta \Phi} \nn \\
&& \quad \quad \quad \quad \times \Big[(\cM^\dagger \cM + \beta_i)^{-1} \xi\Big].
\eea
The quantities $[(\cM^\dagger \cM + \beta_i)^{-1} \xi]$ can be efficiently determined by a multi-shift conjugate gradient (CG) inverter\cite{Jegerlehner:1996pm}.

The Pfaffian phase may be re-incorporated in the expectation value of an observable $\cO$ by re-weighting as
\beq
\langle \cO \rangle = \frac{\sum_i (\cO e^{i \alpha})}{\sum_i (e^{i \alpha})},
\eeq  
where the sum runs over all members of the phase quenched ensemble.

\section{Quantum Gravity Results from the Lattice}
\label{sec:latt-results}

In this section we discuss the results from Monte Carlo simulations of the maximally supersymmetric Yang-Mills theory at finite temperature. The most interesting quantity to compute is the internal energy $E_{\rm gauge}$ of the system. It has the form  
\beq
\label{eq:energy-gauge}
\frac{1}{T^2} E_{\rm gauge} = - \frac{3}{\beta N^2} \left( \langle S_B \rangle - S_0 \right),
\eeq
where $S_0$ denotes the extensive zero-point energy contribution and it is given by 
\beq
S_0 = \frac{9}{2} M (N^2 -1).
\eeq

The other observables of interest are the expectation value of the Polyakov loop
\beq
\langle P \rangle \equiv \frac{1}{N} \Tr U,
\eeq
with the holonomy matrix $U$ defined as 
\beq
U \equiv \cP \exp \left(i \oint^\beta d\tau A_0(\tau) \right),
\eeq
with $\cP$ denoting a path-ordered product and the ``extent of space"
\beq
\langle R^2 \rangle = \left\langle \frac{1}{N \beta} \oint^\beta d\tau~ \Tr (X_i)^2 \right\rangle.
\eeq 

The expectation value of the Polyakov loop plays the role of an order parameter for the confining-deconfining phase transition in the thermal theory. It is non-zero at all temperatures in the supersymmetric theory. This implies that there is no phase transition in the supersymmetric theory as predicted by gauge-gravity correspondence\cite{Barbon:1998cr,Aharony:2005ew}. 

One could also check if the Polyakov loop can be fitted to the form
\beq
\langle P \rangle = \exp \left(- \frac{a}{T} + b\right),
\eeq
which is a characteristic behavior in a deconfined theory. 

The instability related to the flat directions associated with the scalar eigenvalues can be probed by studying the divergence of $\langle R^2 \rangle$. In Ref. [\refcite{Anagnostopoulos:2007fw}] it has been shown, for numerical simulations based on a non-lattice approach, that the the divergence is stabilized as the value of $N$ is increased. This is indeed expected since there is a meta-stable thermal equilibrium in the theory with a decay rate that is very small at large $N$.

The numerical results of the internal energy of the thermal gauge theory are found in Refs. [\refcite{Anagnostopoulos:2007fw,Catterall:2007fp,Hanada:2008gy,Hanada:2008ez,Hanada:2013rga,Catterall:2008yz,Catterall:2009xn,Kadoh:2015mka,Filev:2015hia}].

Recently, the authors of Ref. [\refcite{Filev:2015hia}] computed the internal energy of the gauge theory from Monte Carlo simulations at several lattice spacings and extrapolated the result to the continuum limit. Their simulations show that the Polyakov loop is largely independent on the lattice spacing for all temperatures probed. The extend of space also experienced very weak lattice effects. However, the internal energy is affected by lattice effects. Since the effects of lattice spacing died out linearly in the simulations, they were able to extrapolate the energy to zero lattice spacing. They used the $\alpha'$ corrected expression Eq. \eqref{eq:alpha-prime-corrected} while comparing the lattice results with that of the gravity theory and found good agreement. 

They also observed that the model became unstable while simulating the theory with smaller $N$, which is related to Hawking radiation in the dual gravitational theory\cite{Catterall:2009xn,Hanada:2013rga}. Thus one needs larger matrices for a clear comparison with the AdS/CFT predictions. The results they found were in excellent overall agreement with the studies performed in Refs. [\refcite{Anagnostopoulos:2007fw}] and [\refcite{Kadoh:2015mka}].

In Refs. [\refcite{Kadoh:2014hsa,Kadoh:2015mka}] Kadoh and Kamata simulated the topologically twisted version of the BFSS model. Since they have one additional unintegrated auxiliary field in the theory they used a modified expression for the internal energy 
\beq
\frac{1}{T^2} E_{\rm gauge} = - \frac{3}{\beta N^2} \left( \langle S_B \rangle - \frac{9 + k}{2} M (N^2 -1) \right),
\eeq
where $k$ denotes the number of unintegrated auxiliary fields, and they simulated the theory with $k = 1$. The $Q$-exactness of the lattice action implies that the internal energy vanishes in the zero temperature limit where the effect of the supersymmetry breaking boundary conditions vanishes.

Their results for the internal energy of the gauge theory coincide with the results of the high temperature expansion\cite{Kawahara:2007ib} at higher values of temperatures (See Fig. 3 of Ref. [\refcite{Kadoh:2015mka}].) The lattice data smoothly approach the theoretical prediction of the gravity side as the temperature is decreased, indicating a strong evidence for the gauge-gravity duality. 

Kadoh and Kamata were able to probe the internal energy at the lower temperature region (See Fig. 4 of Ref. [\refcite{Kadoh:2015mka}]). Their investigations show that the lattice data approach the gravity prediction and predict that they are likely to coincide with it as the temperature decreases further. The temperatures they used in the simulations were not low enough to explain the leading behavior on the gravity side, unfortunately. To obtain quantitative results for the leading-order term, simulations at further low temperatures are required. 

Instead, they studied the contribution of the next-to-leading order term by fitting the lattice data using the following formula
\beq
\label{eq:fit-formula}
E(T) = 7.41 ~T^{2.8} + C ~T^p,
\eeq   
where $C$ and $p$ are the fit parameters. From Eq. \eqref{eq:alpha-prime-corrected}, if the duality conjecture is true, the value $p$ obtained from lattice data should be $4.6$ within the statistical errors. The fit performed in Ref. [\refcite{Kadoh:2015mka}] using 5 lattice points within the range $0.375 \leq T \leq 0.475$ gave
\beq
\label{eq:results-C-p}
C = 9.0(2.6),\quad p = 4.74(35).
\eeq
The value of $p$ is consistent with the theoretical prediction from the gravity side within about seven percent statistical error. This is, remarkably, the first lattice result of the NLO term, which quantitatively shows the validity of the gauge-gravity duality in this thermal system.

In Ref. [\refcite{Hanada:2008ez}], the NLO term was estimated from the numerical simulation based on the momentum sharp cut-off method using the same fit formula Eq. \eqref{eq:fit-formula}, in a little higher temperature region $0.5 \leq T \leq 0.7$. They obtain the following values for $C$ and $p$ after the fit  
\beq
C = 5.55(7),\quad p = 4.58(3).
\eeq
They are consistent with the results of Kadoh and Kamata, Eq. \eqref{eq:results-C-p}, within two sigma. 

The results in Ref. [\refcite{Kadoh:2015mka}] were obtained at a finite lattice spacing. It would be interesting to see if the results get better once simulations were performed at different lattice spacings and the results are extrapolated to the continuum limit.

The maximally supersymmetric Yang-Mills quantum mechanics describes the open string degrees of freedom attached to the D0-branes, which are decoupled from the degrees of freedom in the bulk in the decoupling limit given in Eq. \eqref{eq:ft-limit}. The above results imply that we can understand the microscopic origin of the black hole thermodynamics, including the $\alpha'$ corrections, in terms of the open strings attached to the D0-branes. The statement is valid for any value of $\lambda$ and $N$, and lattice is a well suited tool to probe the theory at such arbitrary values of the parameters.

\section{The BMN Matrix Model}
\label{sec:BMN-model}

Recently, a variation of the BFSS matrix model, which has been conjectured to describe a discrete light-cone quantization of M-theory on a pp-wave background has been formulated\cite{Berenstein:2002jq}. This model, known as the BMN model or the Plane Wave Matrix Model, can be constructed as a one-parameter deformation of the BFSS matrix model.  

The dual gravitational geometries to the vacua of the BMN model were constructed in Refs. [\refcite{Lin:2004nb,Lin:2005nh}]. These supersymmetric vacuum geometries, including the one dual to the trivial vacuum, asymptote to the plane wave solution of M-theory.

The BMN model has several advantages over the BFSS model. It has a discrete energy spectrum and a well-defined canonical ensemble. It is convenient to define two dimensionless parameters in this model -- a dimensionless coupling constant 
\beq
g \equiv \frac{\lambda}{\mu^3},
\eeq
with $\lambda$ the 't Hooft parameter and $\mu$ the mass deformation parameter; and 
\beq
t \equiv \frac{T}{\mu},
\eeq
with $T$ the temperature of the system. These two dimensionless quantities can be used to parameterize a two-dimensional phase diagram of the theory. This means that we can use the dual gravitational description at large $N$ and strong coupling $g \gg 1$ to predict several observables as functions of the dimensionless temperature $t$. The BMN model is expected to have a phase transition whose critical temperature should be easy to measure using Monte Carlo simulations.

In BMN model the flat directions associated with the scalar eigenvalues are removed by the mass terms. Recent efforts to study the BFSS matrix model at strong coupling, using both lattice and non-lattice techniques, suffer from the problem of flat directions. The existence of these flat directions means that, at finite temperature, the partition function is formally divergent. It was shown in Ref. [\refcite{Catterall:2009xn}] that when Monte Carlo simulations of this theory are performed, this divergence eventually causes the simulations to break down.

We write down the action of the BMN model 
\bea
\label{eq:BMN-action}
S &=& \frac{N}{2 \lambda} \oint^\beta d\tau~ \Tr \Big( - \sum_i (D_\tau X_i)^2 - \sum_{i < j} [X_i, X_j]^2 + \psi^T \gamma_\tau D_\tau \Psi \nn \\
&&\quad \quad \quad \quad + \sum_i \Psi^T \gamma_i [X_i, \Psi ] - \mu^2 \sum_{i=1}^3 (X_i)^2 - \frac{\mu^2}{4} \sum_{i = 4}^9 (X_i)^2 \nn \\
&&\quad \quad \quad \quad - 2 \sqrt{2} \mu \epsilon_{ijk} X_i X_j X_k + \frac{3}{4} \mu \Psi^T \gamma_{123} \Psi \Big),  
\eea 
where $X_i$ are nine scalars and $\Psi$ represents the fermions. The fields of the theory are in the adjoint representation of $SU(N)$. $D_\tau$ is the gauge covariant derivative and
\beq
\gamma_{123} = \frac{1}{3!} \epsilon_{ijk} \gamma_i \gamma_j \gamma_k.
\eeq

The first four terms of Eq. \eqref{eq:BMN-action} are the same as that of the BFSS matrix model. The rest of the terms give supersymmetry preserving masses to the fields and add a Myers term. As mentioned above, the addition of these terms gives this model technical advantages compared with the BFSS. Firstly, in the limit $\mu \to \infty$ the model becomes weakly coupled and can be studied using perturbative methods\cite{Dasgupta:2002hx}. The addition of these terms lifts the moduli space of the BFSS model giving instead a discrete set of vacua\cite{Berenstein:2002jq}. This is indeed important for numerical simulations as it means the theory is well defined in the canonical ensemble. 

If we consider the scalar fields $X_i$ for $i = 1,2,3$ we can write the potential as 
\beq
V = - \Tr \left[ \left( \mu X_i + \frac{1}{\sqrt{2}} \epsilon_{ijk} [X_i, X_j] \right)^2 \right],
\eeq
which is minimized when the scalar fields $X_i$ are proportional to generators of $SU(2)$. The discrete vacua correspond to the various ways of forming $N \times N$ matrices that generate $SU(2)$, and can be put in one-to-one correspondence with the integer partitions of $N$. It was shown in Ref. [\refcite{Dasgupta:2002ru}] that the vacua are free from quantum corrections and thus persist at strong coupling.  

To write down the lattice action, the parameters of the continuum BMN action Eq. \eqref{eq:BMN-action} are rescaled in the following way 
\bea
X &\to& \mu M X, \\
D_\tau &\to& \mu M D_\tau, \\
\tau &\to& \frac{\tau}{\mu M}, \\
\Psi &\to& (\mu M)^{\frac{3}{2}} \Psi,
\eea
where $M$ is a dimensionless number that will become the number of lattice sites. 

The lattice action is
\bea
\label{eq:BMN-latt-action}
S &=& \frac{\mu^3 M^3 N}{2 \lambda} \oint^{\mu \beta M} d\tau~ \Tr \Big( - \sum_i (D_\tau X_i)^2 - \sum_{i < j} [X_i, X_j]^2 + \psi^T \gamma_\tau D_\tau \Psi \nn \\
&&\quad \quad + \sum_i \Psi^T \gamma_i [X_i, \Psi ] -  \frac{1}{M^2} \sum_{i=1}^3 (X_i)^2 - \frac{1}{4 M^2} \sum_{i = 4}^9 (X_i)^2 \nn \\
&&\quad \quad - 2 \sqrt{2} \frac{\mu}{M} \epsilon_{ijk} X_i X_j X_k + \frac{3}{4 M} \Psi^T \gamma_{123} \Psi \Big). 
\eea 

All the fields in the above action are dimensionless. The next step is to integrate out the fermions to get the following partition function
\beq
Z = \int DA DX~{\rm Pf}(\cM) e^{-S_B},
\eeq
with the bosonic action
\bea
S_B &=& \frac{\mu^3 M^3 N}{2 \lambda} \oint^{\mu \beta M} d\tau~ \Tr \Big( - \sum_i (D_\tau X_i)^2 - \sum_{i < j} [X_i, X_j]^2 - \frac{1}{M^2} \sum_{i=1}^3 (X_i)^2~~~~ \nn \\
&&\quad \quad \quad - \frac{1}{4 M^2} \sum_{i = 4}^9 (X_i)^2 - 2 \sqrt{2} \frac{\mu}{M} \epsilon_{ijk} X_i X_j X_k \Big), 
\eea
and the fermion operator
\beq
\cM = \gamma_\tau D_\tau + \sum_i \gamma_i [X_i, ~\cdot~] + \frac{3}{4M} \gamma_{123}.
\eeq

The lattice theory has an overall coupling
\beq
\kappa = \frac{\mu^4 N \beta}{2 \lambda} M^3,
\eeq
where $\beta$ is the inverse temperature. 

One could then proceed to simulate this model using RHMC algorithm. The continuum limit of the theory is defined by taking the number of lattice points $M \to \infty$ with $\mu^4 N \beta/(2 \lambda)$ fixed. 

In Ref. [\refcite{Catterall:2010gf}] Catterall and van Anders presented the first results of the lattice simulations of the BMN model. They used a naive lattice action, and concentrated on studying the Hagedorn/deconfinement transition in the model. They studied the theory at a fixed temperature, measured in units of the mass deformation $\mu$, as a function of the 't Hooft coupling $\lambda$ (measured in the same units). They used a quenched approximation to explore the dependence of the critical behavior on the rank of the gauge group $N$ and the number of lattice points $M$. The quenched approximation is much less computationally demanding and can be used to estimate reasonable choices of parameters for simulating the full theory. Their simulations indicate that one can get a reasonable approximation of the continuum large $N$ behavior with modest values of both $N$ and size of the lattice. By simulating the model at fixed temperature over a range of coupling they showed that the model exhibits a deconfinement transition when the 't Hooft coupling is of order one. They also found that the critical value of the coupling in the case of quenched simulations was of the same order as in the theory with dynamical fermions. 

According to the conventions of Ref. [\refcite{Costa:2014wya}], Catterall and van Anders simulated the BMN at fixed finite temperature, $T/\mu = 1/3$,  and as they varied the coupling, they observed a first order phase transition for $0.03 \lesssim g \lesssim 0.045$. This result is not in direct contradiction with the results of Ref. [\refcite{Costa:2014wya}] but it implies a non-monotonic behavior of the critical temperature as a function of the coupling $g$, complicating the phase diagram computed in Ref. [\refcite{Costa:2014wya}].

In Ref. [\refcite{Costa:2014wya}] the authors found a Hawking-Page like phase transition in the dual gravitational description of the BMN model and predicted the strong coupling limit of the critical temperature
\beq
\lim_{g \to \infty} \frac{T_c(g)}{\mu} = 0.105905(57).
\eeq  
It would be remarkable to confirm this prediction with Monte Carlo simulations of the BMN model at strong coupling. 

\section{D1-branes and Black Hole - Black String Transitions}
\label{sec:2d-16-theory}

In this section we briefly discuss the lattice results from gauge-gravity duality connecting two-dimensional maximally supersymmetric Yang-Mills theory and type IIB supergravity solution describing the thermal vacuum of black hole carrying electric $D1$-brane charge. 

On the gauge theory side we have a large $N$ finite temperature two-dimensional maximally supersymmetric $SU(N)$ Yang-Mills theory, in the 't Hooft limit, with coupling $\lambda = N g_{YM}^2$, with the spatial direction compactified. In Euclidean time, this implies the Yang-Mills theory is defined on a rectangular 2-torus, with time cycle size $\beta$, and space cycle size $R$. The fermion boundary conditions distinguish the two cycles, being anti-periodic on the time cycle and periodic boundary conditions on the space cycle. 

The action of the Euclidean theory is
\bea
S &=& \frac{N}{\lambda} \int_{T^2} d \tau dx~ \Tr \Big( \qtr F_{\mu\nu}^2 + \hf \sum_i [ D_\mu \phi_i , D_\mu \phi_i ]^2 - \qtr \sum_{i,j} [ \phi_i , \phi_j ]^2 \nn \\
&&\quad \quad \quad \quad \quad \quad + {\rm ~fermions} \Big),
\eea
where $i, j = 1, \ldots, 8$ and $\phi_i$ are the 8 adjoint scalars, and $\tau$ is the coordinate on the time circle, and $x$ the coordinate on the space circle. 

It is convenient introduce two dimensionless couplings in the theory, 
\beq
r_\tau = \lambda^\hf \beta,~~r_x = \lambda^\hf R,
\eeq
which give the dimensionless radii of the time and space circles respectively. 

We will be interested in the expectation values of the Polyakov loops on the time and space circles,
\beq
\langle P_{\tau, x} \rangle = \frac{1}{N} \left\langle \left| \Tr \left(\cP \exp \left[ i \oint A_{\tau, x}\right] \right) \right| \right\rangle.
\eeq
At large $N$, they give order parameters for confinement/deconfinement (or center symmetry breaking) phase transitions in the thermal gauge theory.

There are several interesting limits of this theory, as discussed in Refs. [\refcite{Aharony:2004ig,Aharony:2005ew}]. We are interested in the large torus limit, $1 \ll r_x, r_\tau$ where the string theory dual may be described by a supergravity theory\cite{Itzhaki:1998dd}. Having a supergravity description of the full string theory dual allows certain behaviors of the theory to be studied using simple semi-classical gravity reasoning, which allows powerful predictions to be inferred for the dual SYM.

The dual type IIB string theory is given by the `decoupling limit' of $N$ coincident $D1$-branes\cite{Itzhaki:1998dd}. Since our Euclidean SYM is defined on a torus, the string dual is too, being at finite temperature and having one spatial direction compactified into a circle radius $R$ with periodic fermion boundary conditions.

One finds that for $1 \ll r_\tau \ll r_x^2$ this string theory can be described effectively by its supergravity sector. The type IIB supergravity solution describing the thermal vacuum is a black hole, carrying electric $D1$-brane charge. In the type IIB regime,  $1 \ll r_\tau \ll r_x^2$, we expect $P_\tau \ne 0$ but $P_x = 0$. In the type IIA regime, where $1 \ll r_\tau$ and $r_x^{4/3} \ll r_\tau$, we have $P_\tau \ne 0$, and $P_x \ne 0$ for $r_x^2 \leq c_{\rm crit} r_\tau$ and $P_x = 0$ for $r_x^2 > c_{\rm crit} r_\tau$, with $c_{\rm crit}$ an order one constant with $c_{\rm crit} > 2.29$. We note that in the regime where both type IIA and IIB apply, they give consistent results. Thus in the large torus, supergravity regimes, the SYM is always deconfined in the time direction, and there is a first order deconfinement/confinement transition in the space direction at 
\beq
r_x^2 = c_{\rm crit} r_\tau.
\eeq
In the simplest picture, the above relation characterizes the Gregory-Laflamme (GL) type first order phase transition\cite{Gregory:1993vy} for $1 \ll r_\tau$, with $c_{\rm crit} > 2.29$.

\subsection{Supersymmetric Lattices for 2D SYM}
\label{sec:susy-lattice-2D}

In this section we briefly discuss the manifestly supersymmetric lattice action used to simulate the strongly coupled two-dimensional SYM theory. The lattice theory is based on a twisted version of the four-dimensional maximally supersymmetric theory. The continuum twist of $\cN = 4$ we are interested in was first written down by Marcus\cite{Marcus:1995mq}. As shown by Kaplan and \"Unsal\cite{Kaplan:2005ta, Unsal:2006qp} this twisted version of the theory can be recovered by taking the continuum limit of a lattice theory constructed using the idea of orbifold projection. Catterall\cite{Catterall:2007kn} has constructed a direct lattice discretization of this theory combining the ideas of topological twisting and \KD fermions. These two lattice formulations lead to identical supersymmetric lattices for this theory\cite{Catterall:2009it}. This four-dimensional twisted theory can be most compactly expressed as dimensional reduction of a five-dimensional theory in which the ten bosonic fields (one four-component gauge field and six scalars) are realized as the components of a complexified five-dimensional gauge field  $\cA_m, m = 1 \ldots 5,$ while the sixteen single-component twisted fermions are realized as the sixteen components of a \KD field $\{\eta, \psi_m, \chi_{mn}\}$. Twisting leads to a scalar fermion in the theory -- it implies the existence of a nilpotent supersymmetry which will be preserved in the lattice theory. 

The scalar supersymmetry charge acts on the continuum fields in the following way
\bea
\cQ \cA_m &=& \psi_m \\
\cQ \psi_m &=& 0 \\
\cQ \cAb_m &=& 0 \\
\cQ \chi_{mn} &=& -\cFb_{mn} \\
\cQ \eta &=& d \\
\cQ d &=& 0
\label{Qsusy}
\eea
The field $\cA_m$ is the complex conjugate of $\cA_m$ and $\cFb_{mn}$ is the complexified curvature made out of $\cAb_m$. The scalar field $d$ is an auxiliary field that is included to close the $\cQ$ supersymmetry algebra and is subsequently integrated out of the final lattice action. 

The action of the twisted theory can be written as the sum of two terms -- a $\cQ$-exact piece and a $\cQ$-closed term.

The $\cQ$-exact piece has the form
\beq
S = \frac{1}{g_{YM}^2} \cQ \int \Tr \left( \chi_{mn} \cF_{mn} + \eta [ \cDb_m, \cD_m ] - \hf \eta d \right),
\label{2d-action-exact}
\eeq
and the $\cQ$-closed term 
\beq
S_{\rm closed} = - \frac{1}{8g_{YM}^2} \int \Tr \epsilon_{mnpqr} \chi_{qr} \cDb_p \chi_{mn}.
\label{closed}
\eeq

The $\cQ$-exact piece is trivially $\cQ$ supersymmetry invariant. The supersymmetric invariance of the $\cQ$-closed term can be shown using the Bianchi identity for the complexified covariant derivatives 
\beq
\epsilon_{mnpqr} \cDb_p \cFb_{qr} = 0.
\eeq 

We can discretize the theory in a straightforward manner. The complex continuum gauge fields are represented as complexified Wilson gauge links 
\beq
\cU_\mu(\bn) = e^{\cA_\mu(\bn)}
\eeq
living on positively oriented links $\hatbe_\mu, \mu = 1, \ldots, 4,$ of a four-dimensional hypercubic lattice with integer sites $\bn$. The field $\cU_5$ is placed on the body diagonal of the hypercube corresponding to a relative position vector 
\beq
\hatbe_5 = (-1, -1, -1, -1).
\eeq
Notice that we have the restriction 
\beq
\sum_{m=1}^5 \hatbe_m = 0
\eeq
in order for the action to be gauge-invariant on the lattice. These fields transform in the usual way under the $U(N)$ lattice gauge transformations. For example, we have the following gauge transformation rule for the complexified gauge link
\beq
\cU_m(\bn) \to G(\bn) \cU_m(\bn) G^\dagger(\bn + \hatbe_m),
\eeq
for $G \in U(N)$.

Supersymmetric invariance then implies that $\psi_m(\bn)$ live on the corresponding links and transform identically to $\cU_m(\bn)$. The scalar fermion $\eta(\bn)$ is associated with a site and transforms like a site field
\beq 
\eta(\bn) \to G(\bn) \eta(\bn) G^\dagger(\bn).
\eeq

The field $\chi_{mn}$ is placed on the diagonal links connecting sites $\bn + \hatbe_m + \hatbe_n$ and $\bn$. This choice of orientation will again be necessary to ensure gauge-invariance. The scalar lattice supersymmetry transformation is identical to that in the continuum after the replacement
\beq
\cA_m \to \cU_m.
\eeq
Most importantly it remains nilpotent, which means that we can guarantee invariance of the $\cQ$-exact part of the lattice action by replacing the continuum fields by their lattice counterparts. 

We also need a prescription for replacing continuum derivative operators by gauge-covariant finite difference operators. We use the following prescriptions\cite{Catterall:2007kn,Damgaard:2007be,Damgaard:2008pa}. A forward lattice covariant difference operator is used to denote a curl-like operation on the twisted fields 

\beq
\cD^{(+)}_m f_n(\bn) = \cU_m(\bn) f_n(\bn + \hatbe_m) - f_n(\bn) \cU_m(\bn + \hatbe_n), 
\eeq
while backward lattice covariant difference operator is used to denote a divergence-like operation
\beq
\cDb^{(-)}_m f_m(\bn) = f_m(\bn) \cUb_m(\bn) - \cUb_m(\bn - \hatbe_m) f_m(\bn - \hatbe_m).
\eeq

They reduce to the usual adjoint covariant derivatives in the naive continuum limit. It is also guaranteed that the resultant expressions transform covariantly under lattice gauge transformations.

It is possible to discretize Eq. \eqref{closed} in such a way that it is indeed exactly invariant under the twisted supersymmetry
\beq
S_{\rm closed} = - \frac{1}{8g_{YM}^2} \sum_{\bn} \Tr \epsilon_{mnpqr} \chi_{qr} (\bn + \hatbe_m + \hatbe_n + \hatbe_p) \cDb^{(-)}_p \chi_{mn}(\bn + \hatbe_p),
\eeq
and can be seen to be supersymmetric since the lattice field strength satisfies an exact Bianchi identity
\beq
\epsilon_{mnpqr} \cDb^{(+)}_p \cFb_{qr} = 0
\eeq

Assembling all these together we arrive at the supersymmetric lattice action (See Refs. [\refcite{Catterall:2009it,Joseph:2011xy}] for reviews)
\bea
S &=& S_{\rm closed} + \frac{1}{g_{YM}^2} \sum_{\bn} \Tr \Big( \cF^{\dagger}_{mn}(\bn) \cF_{mn}(\bn) + \hf \left( \cDb^{(-)}_m \cU_m(\bn) \right)^2 \nn \\
&&\quad \quad \quad \quad \quad - \chi_{mn}(\bn) \cD^{(+)}_{\left[m\right.} \psi_{\left.n\right]}(\bn) - \eta(\bn) \cDb^{(-)}_m \psi_m(\bn) \Big),
\label{4action}
\eea
where we have taken the $\cQ$-variation and the auxiliary field has been integrated out. This action is gauge-invariant, free of doublers and possesses the one exact supersymmetry charge $\cQ$.

In order to arrive at the two-dimensional theory we perform a simple dimensional reduction along two lattice directions using periodic boundary conditions. The resultant lattice action corresponds in the naive continuum limit to the target sixteen supercharge Yang-Mills theory in two dimensions. In this limit its exact supersymmetry is enhanced to four supercharges corresponding to the four scalar fermions that now appear in the dimensionally reduced theory\cite{Catterall:2009it}.

We will be interested in this theory at the large $N$ limit with 't Hooft coupling $\lambda$. The lattice theory is then governed by the coupling 
\beq
\kappa = \frac{N L M}{2 r_\tau^2},
\eeq
where $L$ and $M$ denote the number of lattice sites in the spatial and temporal directions, respectively. 

We use periodic boundary conditions for the fields on the remaining spatial circle and anti-periodic boundary conditions for fermions in the temporal direction in order to access the theory at finite temperature. Simulations can be carried out using RHMC algorithm. It has been shown that the existence of a non-compact moduli space in the theory may render the thermal partition function divergent\cite{Catterall:2009xn}. In order to regulate this divergence one introduces mass terms for the scalar fields appearing in the lattice action with a dimensionless mass parameter
\beq
m = m_{\rm phys} \beta.
\eeq

The mass terms added to the action have the form
\beq
S_M = \frac{m^2}{g_{YM}^2} \sum_\bn \Tr \left( \cU_\mu^\dagger\cU_\mu + \left(\cU_\mu^\dagger\cU_\mu\right)^{-1} - 2 \right).
\eeq

Such mass terms are effective at suppressing arbitrarily large fluctuations of the exponentiated scalar fields and reduces to simple mass terms for small fluctuations characterizing the continuum limit. These infrared regulator terms break supersymmetry softly and lift the quantum moduli space of the theory. In Ref. [\refcite{Catterall:2010fx,Catterall:2010ya}] simulations were performed for a range of the mass parameter in order to allow for an extrapolation $m \to 0$. 

\subsection{Results on Gregory-Laflamme Phase Transition}
\label{sec:results-GL}

In this section we present the numerical results of the studies performed in Refs. [\refcite{Catterall:2010fx,Catterall:2010ya}]. The simulations are based on the Polyakov lines for both the thermal and spatial circle. These are defined in the usual way
\bea
P_x &=& \frac{1}{N} \left\langle \left| \Tr \Pi_{a_x=0}^{L-1} U_{a_x} \right| \right\rangle, \\
P_\tau &=& \frac{1}{N} \left\langle \left| \Tr \Pi_{a_\tau=0}^{T-1} U_{a_\tau} \right| \right\rangle,
\eea
where the unitary piece of the complexified link $\cU_\mu$ has been extracted to compute these expressions. The spatial and temporal Polyakov lines are evaluated as functions of $r_\tau$ for various lattice sizes and number of colors.

The temporal Polyakov line remained close to unity in the simulations over a wide range of $r_\tau$, indicating that the theory is (temporally) deconfined. The spatial Polyakov line has a different behavior, they take values close to unity for small $r_\tau$ and fall rapidly to plateau at much smaller values for large $r_\tau$. It is tempting to see the rather rapid crossover around $r_\tau \sim 0.2$ as a signal for a would be thermal phase transition as the number of colors is increased.  This conjecture is seen to be consistent with the lattice data. The plateau evident at large $r_\tau$ falls with increasing $N$ and the crossover sharpens. This is consistent with the thermal system developing a sharp phase transition in the large $N$ limit. 

Simulations show that the $m = 0$ model does exhibit the same thermal instability observed in the case of sixteen supercharge quantum mechanics for sufficiently low temperature $r_\tau >> 1$, in agreement with the general arguments given in Ref. [\refcite{Catterall:2009xn}].

The gravity prediction for the parametric behavior $r_x^2 = c_{\rm crit} r_\tau$ is consistent with the lattice data, and $c_{\rm crit}$ was estimated to be\cite{Catterall:2010fx} 
\beq
c_{\rm crit} \simeq 3.5,
\eeq
which indeed obeys the gravity prediction that $c_{\rm crit}$ is order one and $c_{\rm crit} > 2.29$. 

The value of the ratio 
\beq
\alpha \equiv \frac{c_{\rm crit}}{2.29}
\eeq
gives the ratio of the GL thermal phase transition temperature to the GL dynamical instability temperature (the minimum temperature to which uniform strings can be supercooled), so 
\beq
\alpha = \frac{T_{\rm GL~phase}}{T_{\rm GL~instab}}.
\eeq

The GL instability temperature is known\cite{Aharony:2004ig} (corresponding to the behavior $r_x^2 = 2.29 r_\tau$ at strong coupling) however, the GL phase transition temperature is not known in the gravity theory as the localized solutions have not been constructed yet.
 
The lattice estimation\cite{Catterall:2010fx} 
\beq
\alpha \simeq 1.5
\eeq
provides a prediction for the thermal behavior of the gravity solutions. This was the first time a prediction about the properties of non-trivial classical gravity solutions had been made from the Yang-Mills side of a gauge-gravity correspondence. Since the dual localized black hole solutions have not been constructed, this constitutes a prediction for these non-trivial gravity solutions, which hopefully will be tested by their construction in the near future.

\section{Future Directions}
\label{sec:future-directions}

In this review we have focused on maximally supersymmetric lattice Yang-Mills theories in one and two dimensions and showed how they can validate and provide insight into the gauge-gravity duality conjecture. For the case of one-dimensional theories we have detailed the duality conjecture connecting D0-branes and black hole geometries at finite temperature to maximally supersymmetric thermal gauge theories at the same temperature. The internal energy of a quantum black hole with its leading order quantum gravity and string theory corrections has been derived recently in the gravitational theory. The predictions from supergravity can be verified by simulating the dual strongly coupled thermal gauge theories on the lattice and calculating the internal energy of the thermal system for a given number of colors and 't Hooft coupling. Recent simulations indicate that the gauge theory results give excellent agreement with the quantum gravity predictions, thus proving strong evidence for the validity of the duality conjecture. We detailed the naive and manifestly supersymmetric constructions of maximally supersymmetric lattice Yang-Mills theories and the simulation algorithms used to probe the strong coupling regimes of these gauge theories at finite temperature and number of colors. We have also briefly detailed the results on the duality in two-dimensional sixteen supercharge thermal gauge theories and corresponding supergravity solutions. We also note that the existing results for the two-dimensional system can be improved using improved lattice simulation methods recently proposed in Refs. [\refcite{Schaich:2014pda,Catterall:2015ira}].  

It would be interesting to extend the analysis of the gauge-gravity duality to three- and four-dimensional Yang-Mills systems, which are thought to be dual to D2- and D3-branes systems. Such theories are easily accessible through the exact lattice supersymmetry formulation.

Another interesting matrix model to investigate in the context of gauge-gravity duality is the Berkooz-Douglas matrix model\cite{Berkooz:1996is}. This model is holographically dual to the back-reacted D0-D4 system. Investigations on this model can give more extensive tests of the gauge-gravity duality. 

Numerical simulations of the sixteen supercharge quantum mechanics model become increasingly difficult for smaller $N$ and lower temperatures. The simulations suffer from technical difficulties related to the divergence of the eigenvalues of the bosonic matrices while sampling the important field configurations that contribute to the action. It would be remarkable if we could tackle this problem using new numerical or analytical methods to provide more insight into the gauge-gravity duality and the internal structure of quantum black holes. 

\section{Acknowledgments}

I thank useful discussions with Simon Catterall, Poul Damgaard, Tom DeGrand, Eric Dzienkowski, Richard Galvez, Joel Giedt, Issaku Kanamori, David Kaplan, Piotr Korcyl, So Matsuura, Dhagash Mehta, David Schaich, Mithat \"Unsal, Aarti Veernala, Robert Wells and Toby Wiseman on lattice supersymmetry and related ideas over the years. This work was supported in part by the German Research Foundation (DFG) and the European Research Council under the European Union's Seventh Framework Programme (FP7/2007-2013), ERC grant agreement STG 279943, “Strongly Coupled Systems”.


\end{document}